# Exploring the Band Structure of Wurtzite InAs Nanowires Using Photocurrent Spectroscopy

*Seyyedesadaf Pournia, Samuel Linser, Giriraj Jnawali, Howard E. Jackson and Leigh M. Smith*
Department of Physics, University of Cincinnati, Cincinnati, OH 45221-0011

*Amira Ameruddin, Philippe Caroff, Jennifer Wong-Leung, Hark Hoe Tan, and Chennupati Jagadish*
Department of Electronic Materials Engineering, Research School of Physics, The Australian National University, Canberra, ACT 2601, Australia

*Hannah J. Joyce*
Department of Engineering, University of Cambridge, 9 JJ Thomson Avenue, Cambridge CB3 0FA, United Kingdom

Address correspondence to leigh.smith@uc.edu

ABSTRACT:

We use polarized photocurrent spectroscopy in a nanowire device to investigate the band structure of hexagonal Wurtzite InAs. Signatures of optical transitions between four valence bands and two conduction bands are observed which are consistent with the symmetries expected from group theory. The ground state transition energy identified from photocurrent spectra is seen to be consistent with photoluminescence emitted from a cluster of nanowires from the same growth substrate. From the energies of the observed bands we determine the spin orbit and crystal field energies in Wurtzite InAs. This information is essential to the development of crystal phase engineering of this important III-V semiconductor.

1. Introduction

III-V semiconductor nanowires (NWs) are quasi-one-dimensional materials which show great promise as a nanoscale platform for efficient and high-speed electronic devices, nanosensors and photovoltaics[1,2]. Among the III-V NWs, InAs is of particular interest because it exhibits a high electron mobility, a low effective electron mass, a large spin orbit energy and a small energy band gap. The spin-orbit energy is nearly as large as the band gap energy resulting in strong momentum-spin coupling which has been utilized in the search for Marjarona fermions[3]. Like many III-V NWs, InAs occurs in both the usual Zincblende (ZB) cubic phase, but also hexagonal Wurtzite (WZ). The hexagonal Wurtzite crystal structure has a lower symmetry than the cubic Zincblende structure which has a large impact on both the band structure and selection rules for optical transitions. The resulting differences in the band structure have been extensively explored for both InP [4,5,6,7] and GaAs [8,9] NWs. This detailed understanding has resulted in the rapid development of ZB/WZ nanowire axial and radial heterostructures in these materials to control the thermal conductivity [10], g-factor and diamagnetic coefficients [11,12,13], and enhanced control of emission quantum efficiencies and detector sensitivities [14,15,16,17,18]. There is also intense interest in developing hexagonal Si, Ge and SiGe alloys as direct-gap materials forming the basis for silicon-based optoelectronics [19,20,21] .

Much less is known about Wurtzite InAs. Experimental measurements of the WZ InAs fundamental gap range from 0.43 eV to 0.54 eV [22, 23, 24,25] and theoretical calculations of the gap range from 0.46 to 0.481 eV [26,27,28,29,30]. There are no direct measures of the valence and conduction band structure, but a number of theoretical calculations exist [26,27,28 ,29 ,30]. In this letter, we use polarized photocurrent spectroscopy in a WZ NW device to determine its band structure over a wide range of energies from 0.3 to 1.2 eV.

2. **Material and Methods**

Wurtzite InAs nanowires were fabricated on a [111]-oriented InAs substrate using the metal-catalyzed MOCVD (metal-organic chemical vapor deposition) growth method with 50 nm gold nanoparticles. A substrate temperature of 500°C and an arsine/trimethylindium ratio of 2.9 were used to achieve WZ crystal growth [31]. Fig. 1 (a) and (b) show a SEM image of the growth substrate and two plan view HRTEM images of a typical nanowire demonstrating the single phase WZ nature of the nanowires. The NWs were removed mechanically from the growth substrate into a methanol solution and dispersed on a p-doped silicon substrate with a 300 μm $SiO_2$ layer on the surface. The dispersed wires were usually small (< 5) bundles of wires that formed through van der Waals forces. A well-separated bundle of wires was identified using optical microscopy for device fabrication. To fabricate the device, two 20 nm Titanium and 300 nm Aluminum contacts were deposited on either end of the wire using photolithography, metal deposition and liftoff. After confirmation that the contacts were conductive using a probe station, the nanowire device was placed into a gold-plated chip carrier which was mounted onto the cold finger of an optical cryostat for low temperature measurements.

Figure 2 shows a typical I-V measurement on a 50 nm nanowire device at room temperature, in the dark and under white light illumination. From this measurement, we estimate the resistivity of the nanowire to be 0.9 Ω.mm. Assuming an electron mobility of 210 $cm^2.V^{-1}.s^{-1}$ (measured in similarly synthesized samples [32]), one can estimate that the carrier density in the nanowire to be n~$3x10^{17}$ $cm^{-3}$ which is also consistent with Ref. [33]. Photocurrent spectra were obtained by fixing the current in the device and measuring the change in the bias voltage as the laser is tuned over a wavelength range. The laser is mechanically chopped and the change in the bias voltage is measured using a lock-in amplifier.

3. Background:

For ZB NWs, optical transitions which promote electrons from the valence bands to the conduction bands are allowed for any polarization of light. The selection rules for optical transitions between the various valence bands and conduction bands in WZ NWs are more complicated. WZ InAs belongs to the $C_{6v}$ point-group symmetry (cubic Zincblende InAs has $T_d$ symmetry). The valence bands at the center of the Brillouin zone can be ordered from highest energy to lowest energy as first (A), second (B), third (C), and fourth (D) valence bands with $\Gamma_9$, $\Gamma_7$, $\Gamma_7$, and $\Gamma_9$ symmetries, respectively (see Fig. 1). Similarly, the lowest lying conduction band ($CB_1$) has $\Gamma_7$ symmetry, and a second conduction band ($CB_2$) at higher energies with $\Gamma_8$ symmetry which results from zone folding of the L-valley in the cubic Zincblende structure to the center of the Brillouin zone (k = 0) in the Wurtzite phase [26].

To determine the optical selection rules for the $C_{6v}$ symmetry group, the light polarized perpendicular (E⊥c) and parallel (E‖c) to the long axis of the nanowire (the Wurtzite c-axis) are associated with $\Gamma_5$ and $\Gamma_1$ symmetries, respectively[34]. An optical transition is allowed for a particular polarization if the product of the initial and final state symmetries contains the symmetry associated with a particular polarization. In equation (1) below, the various band symmetries are multiplied, and the results tabulated.

$$\Gamma_9(A,D) \times \Gamma_7(CB1) = \Gamma_5(E\perp c) + \Gamma_6$$
$$\Gamma_7(B,C) \times \Gamma_7(CB1) = \Gamma_1(E\|c) + \Gamma_6 + \Gamma_5(E\perp c) \quad (1)$$
$$\Gamma_9(A,D) \times \Gamma_8(CB2) = \Gamma_5(E\perp c) + \Gamma_6$$
$$\Gamma_7(B,C) \times \Gamma_8(CB2) = \Gamma_3 + \Gamma_4 + \Gamma_6$$

From equation (1), we see that for hexagonal Wurtzite InAs, optical transitions from the A or D valence bands to $CB_1$ are only allowed for perpendicular polarization, while optical transitions from the B or C valence bands to $CB_1$ are allowed for both polarizations. Perpendicularly polarized light can excite carriers from the A or D valence bands to the second conduction band, however, excitations from either B or C valence bands to $CB_2$ are not allowed at all. Figure 1 provides a summary of all the allowed optical transitions in the $C_{6v}$ symmetry group.

4. **Experiment and Results:**

In this letter we explore the energy band structure and optical selection rules in a WZ InAs NW device by using polarized photocurrent (PC) spectroscopy in the infrared energy range 0.3-1.2 eV (or 4000 to 1000 nm) at 300 K and 10 K. The signal and idler output from an OPO pumped by a 4 W Ti-sapphire laser is continuously tuned from 0.3 to 1.2 eV. The polarization is made linear by using a wire grid polarizer and then rotated using a CaF$_2$ double Fresnel Rhomb rotator to align parallel or perpendicular to the nanowire device. The laser beam was attenuated to an average power of 200 uW and focused onto the sample using a 40X reflective objective to a 1.3-5 micron spot varying with the wavelength of the laser. Representative PC spectra are shown at both 10 K and 300 K in Fig. 3. Errors associated with each transition are estimated by at least two or more measurements taken at different times on the same device as shown in the Supplemental Information. For the PC spectrum taken at 10K, we see that there is no photocurrent at low energies and a clear onset at $0.473 \pm 0.002\ eV$ eV for light polarized perpendicularly to the NW long axis (c-axis), with a peak appearing at $0.482 \pm 0.006\ eV$ signaling the transition from the A valence band to CB$_1$.

For comparison, photoluminescence (PL) measurements were taken on clusters of wires dispersed onto a silicon substrate from the same growth by using 800 nm excitation as an excitation source. The laser was chopped at 300 Hz and focused onto the nanowire cluster using a 40X/0.5 NA reflective objective. PL emitted by the NW cluster was collected by the same objective and focused onto the entrance slit of a 0.2 m spectrometer and dispersed by a 600 line/mm grating. The PL was detected by a lock-in and an InSb pn diode cooled to 77 K. Below the 10 K PC spectra, we display 10 K PL spectrum taken from a cluster of approximately ten similar nanowires, which confirms the assignment of the onset and peak from the PC spectrum. The fundamental band gap for Wurtzite material is thus 60 meV higher than Zincblende InAs band gap as expected from both theory and recent experiments,[22,23,24,25,29,30].

The photocurrent spectrum for light polarized parallel to the NW (c axis) shows a higher energy onset and peaks at $0.550 \pm 0.005\ eV$ which is assumed to be the excitation from the B valence band to $CB_1$. The peak at $0.915 \pm 0.005\ eV$ which is visible for both parallel and perpendicular polarizations is assigned to the transition from the third valence band (C) to the first conduction band ($CB_1$).

At higher energies, two additional but weak peaks are observed only for perpendicular excitation at $0.988 \pm 0.003\ eV$ and $1.065 \pm 0.004\ eV$. Based on the selection rules described previously, we find these transitions can result from transitions either from A to $CB_2$, or D to $CB_1$. In order to most closely match De and Pryor's calculation for the splitting between $CB_1$ and $CB_2$ and the splitting between C and D, we assign the 1.065 eV transition to A to $CB_2$ and the 0.988 eV transition to D to $CB_1$[26]. From this assignment, we find the splitting between $CB_2$ and $CB_1$ to be $0.583 \pm 0.010\ eV$ (De and Pryor predict a splitting of 0.74 eV). We find the splitting between the D and C valence bands to be $0.073 \pm 0.008\ eV$ (De and Pryor predict 0.183 eV). The energy gap and transition energies and energy splittings measured in our experiment are shown in Table 1 and compared with both theoretical calculations and experiments.

Photocurrent spectra taken at 300K (see Fig. 3) shows the first three transitions obeying the expected selection rules, but shifted towards lower energy because of the temperature-dependent shift of the energy gap [35,36,37]. The observed energy splitting between valence bands exhibits no obvious temperature dependence between the 10 K and 300 K measurements, consistent with results in the other material [38].

Using the AB and AC splittings obtained from these measurements, it is possible to extract the spin-orbit energy and crystal field energy for this structure using the quasi-cubic approximation [26,39] shown in equation 2, below:

$$\Delta_{SO,CF} = \frac{1}{2} \left( \Delta_{AB} + \Delta_{AC} \pm \sqrt{\Delta_{AB}^2 + \Delta_{AC}^2 - 4\Delta_{AB}\Delta_{AC}} \right) \qquad (2)$$

As noted in several publications equation 2 results in two energies which may be assigned to either the SO or the CF energies [6, 26, 34]. The spin-orbit energy results from the constituent In and As atoms in these materials so one would expect very similar spin-orbit energies for either Wurtzite or Zincblende InAs whose spin-orbit energy is 0.38 eV [40,41]. Equation 2 provides the two solutions 0.387 eV and 0.113 eV, for the crystal splitting or spin-orbit splitting. It seems reasonable to assign 0.387 eV to the WZ InAs spin orbit energy and 0.113 eV to the crystal field energy. These results are tabulated in Table 1 for both 300 K and 10 K measurements.

**TABLE 1: The energy bands splitting from the experiments at T=10 K and T=300 K**

|  | Wurtzite InAs | | | | | | | $\Delta(E_g^{WZ} - E_g^{ZB})$ (eV) |
|---|---|---|---|---|---|---|---|---|
|  | $E_g$(eV) | $\Delta_{AB}$(eV) | $\Delta_{AC}$(eV) | $\Delta_{AD}$(eV) | $\Delta_{CB1-CB2}$(eV) | $\Delta_{SO}$(eV) | $\Delta_{CF}$(eV) |  |
| T=300K | 0.438 | 0.060 | 0.426 | - | - | 0.387 | 0.098 | 0.7 |
| T=10K | 0.482 | 0.068 | 0.433 | 0.505 | 0.583 | 0.387 | 0.113 | 0.6 |

## 5. Discussion

We now compare the results from our measurements to both theoretical calculations of the WZ InAs band structure, and other experimental measurements of the fundamental gap. Results from the 10K data is compared with the other theoretical and experimental results in Table 2. De and Pryor calculated the WZ band structure for InAs using empirical pseudopotentials [26]. Junior et al calculated the WZ InAs band structure using Density Functional Theorey (DFT) with a modified Becke-Johnson (mBJ) exchange potential with Local Density Approximation (LDA) correlations [27]. Gimitra and Fabian use DFT with semilocal modified Becke-Johnson exchange potentials (TB-mBJ) and with LDA correlations [28].

Bechstedt and Belabbes used DFT with LDA exchange and correlations to calculate the band structure of the various InAs polytypes, including pure WZ [30]. Zanolli et al used DFT calculations using LDA exchange and correlations followed by GW corrections [29].

The parameters for the various AB and AC valence band splittings, and splittings between the two conduction bands, along with SO and CF energies are tabulated from these different calculations in Table 2 and compared with the results here. While all of the theoretical calculations of the fundamental gap agree fairly closely with each other and also with our present measurement, there is substantial variability for the other parameters. For example, theoretical estimates of the AB valence band splitting range from 59 to 105 meV as compared with our measurement of 70 meV [26,27,28]. The theoretical AC valence band splitting ranges from 350 to 470 meV as compared with our measurement of 440 meV [26,27,28]. The theoretical crystal field energies range from 95 to 195 meV as compared with our measurement of 120 meV [26,30]. The theoretical spin orbit energies range from 356 to 379 meV while our measurement is 390 meV [26 ,30]. The largest discrepancy is seen in our measurement of the CB splittings to be 590 meV with theoretical estimates of the CB splitting substantially higher at 706 to 741 meV [26,28].

**TABLE 2: Comparison between the experiment results at T= 10 K and other experimental and theoretical studies**

|  | Theory | Other experiments | PC measurement in figure 3 (10K) |
|---|---|---|---|
| $E_g$ (eV) | $0.47^{29}$ $0.46^{28}$ $0.481^{30,26}$ $0.467^{27}$ | $0.52(7K)^{23}$ $0.477(11K)^{22}$ $0.458(5K)^{24}$ $0.54(5K)^{25}$ $0.43(10K)^{42}$ | 0.482 |
| $\Delta_{AB}$(eV) | $0.066^{28}$ $0.105^{26}$ $0.0592^{27}$ |  | 0.068 |
| $\Delta_{AC}$(eV) | $0.36^{28}$ $0.469^{26}$ $0.3527^{27}$ |  | 0.433 |
| $\Delta_{CB1-CB2}$(eV) | $0.706^{28}$ $0.741^{26}$ |  | 0.583 |
| $\Delta_{CF}$(eV) | $0.095^{30}$ $0.195^{26}$ |  | 0.113 |
| $\Delta_{SO}$(eV) | $0.356^{30}$ $0.379^{26}$ |  | 0.387 |

In order to gauge the strength of the polarization selection rules we now look at the degree of polarization for these different transitions. We define the degree of polarization as DOP = $(I_\perp - I_{\parallel})/(I_\perp + I_{\parallel})$ to describe the polarization dependence of the photocurrent spectra. Based on classical electrodynamics, if a cylinder with dielectric constant $\varepsilon$, is placed in an electric field E, the component of the internal field parallel to the axis of cylinder is the same as the outside field ($E_{\parallel} = E_{0\parallel}$), but it is attenuated in the perpendicular direction ($E_\perp = \frac{2}{1+\varepsilon} E_{0\perp}$) [43]. Since the absorbed energy is $\propto E^2$, for an InAs wire with $\varepsilon$=16.78, for light polarized perpendicular to the NW with an intensity of $P_0$ the intensity of light *inside* the nanowire would be $P_0/79$. Similarly, for light polarized parallel to the NW with an intensity $P_0$ the intensity *inside* the nanowire would also be $P_0$. The degree of polarization in our nanowire is plotted in Fig. 4 for the 10 K and 300 K measurements. A sharp peak centered at 0.48 eV is observed, showing an 85% perpendicularly polarized photocurrent around the fundamental band gap energy.

The response of the device to polarized light is a rather complicated combination of the dielectric response of the nanowire itself, and also the optical selection rules. We attempt to disentangle these responses by using the dielectric response derived above. Assuming that the incident power on the nanowire is the same for parallel and perpendicular polarizations, this can be expressed as [44]:

$$\frac{I\perp}{I\|} = \left(\frac{2}{1+\varepsilon\perp}\right)^2 \frac{\alpha\perp}{\alpha\|} \quad (3)$$

Where $\varepsilon\perp$ is the real part of the dielectric constant in the direction perpendicular to the c axis of wire and $\alpha\perp$ is the absorption constant. Since the $\alpha\perp$ is directly proportional to the imaginary part of the dielectric constant ($\varepsilon''$) [45] and inversely proportional to the index of refraction, and we know that $\varepsilon''$ is proportional to the oscillator strength (f), we find the perpendicular to parallel current ratio to be equal to:

$$\frac{I\perp}{I\|} = \left(\frac{2}{1+\varepsilon\perp}\right)^2 \frac{f\perp}{f\|} \sqrt{\frac{\varepsilon\|}{\varepsilon\perp}} \quad (4)$$

From our measurement, the ratio of the polarized currents is ~12.5 at the A→CB1 transition. By using the dielectric constants for light parallel and perpendicular to the c-axis calculated by De and Pryor ($\frac{\varepsilon\|}{\varepsilon\perp}$~ 0.81) [46], we find $\frac{f\perp}{f\|}$~1100, which shows that excitation of a dipole perpendicular to the nanowire has a much stronger oscillator strength than parallel. This

implies that the absorption of parallel polarized light for this lowest energy transition is nearly negligible as anticipated from group theory (shown previously) [47,48].

6. **Summary and Conclusions**

Using polarized PC spectroscopy over the energy range 0.3 to 1.2 eV, we have identified the energies of four valence and two conduction bands in Wurtzite InAs nanowires. The results agree well with predicted energy gaps from empirical pseudopotential calculations and the polarization of the peaks are in excellent agreement with the optical selection rules for a hexagonal material. From the AB and AC valence band splittings, we extract a spin-orbit energy of 0.38 eV and a crystal field energy of 0.12 eV for Wurtzite InAs. The splitting between the first and second conduction bands is observed to be 0.59 eV. From the degree of polarization measured from the A to CB1 ground state transition we show that the ratio of the perpendicular to parallel oscillator strength is ~1100. Using these band energies, it should now be possible to optimize ab initio band structure calculations. Detailed understanding of the WZ InAs band structure will be essential to the development of crystal phase engineering of this important infrared optoelectronic and high speed electronic materials system.

**Acknowledgements**

We acknowledge the financial support of the NSF through Grants DMR 1507844, DMR 1531373, and ECCS 1509706 and also the financial support of the Australian Research Council and the European Research Council (Grant No. 716471, ACrossWire).. The Australian National Fabrication Facility (ACT Node) is acknowledged for access to the growth facility used in this work. The Australian Microscopy and Microanalysis Research Facility is acknowledged for access to the electron microscopes used in this work.

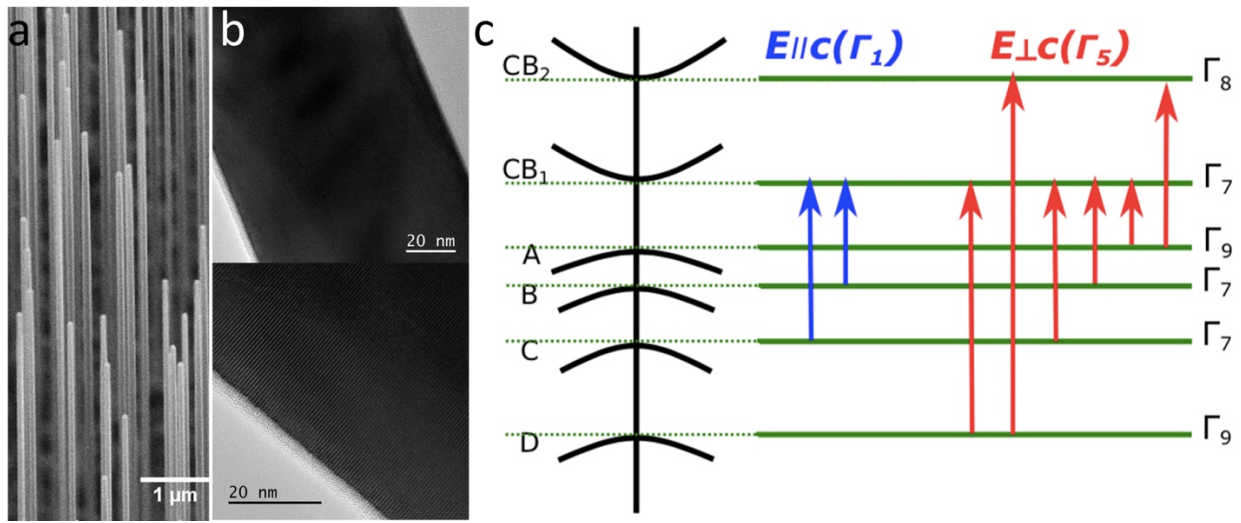

*Figure 1.(a) The SEM image and (b) the TEM of 50 nm diameter Wurtzite InAs nanowires on the initial substrate. (c) shows the band diagram of Wurtzite structure and the allowed optical transitions from four of the valence band (A, B, C, D) and The two conduction* **bands (CB1 and CB2) considering their symmetries.**

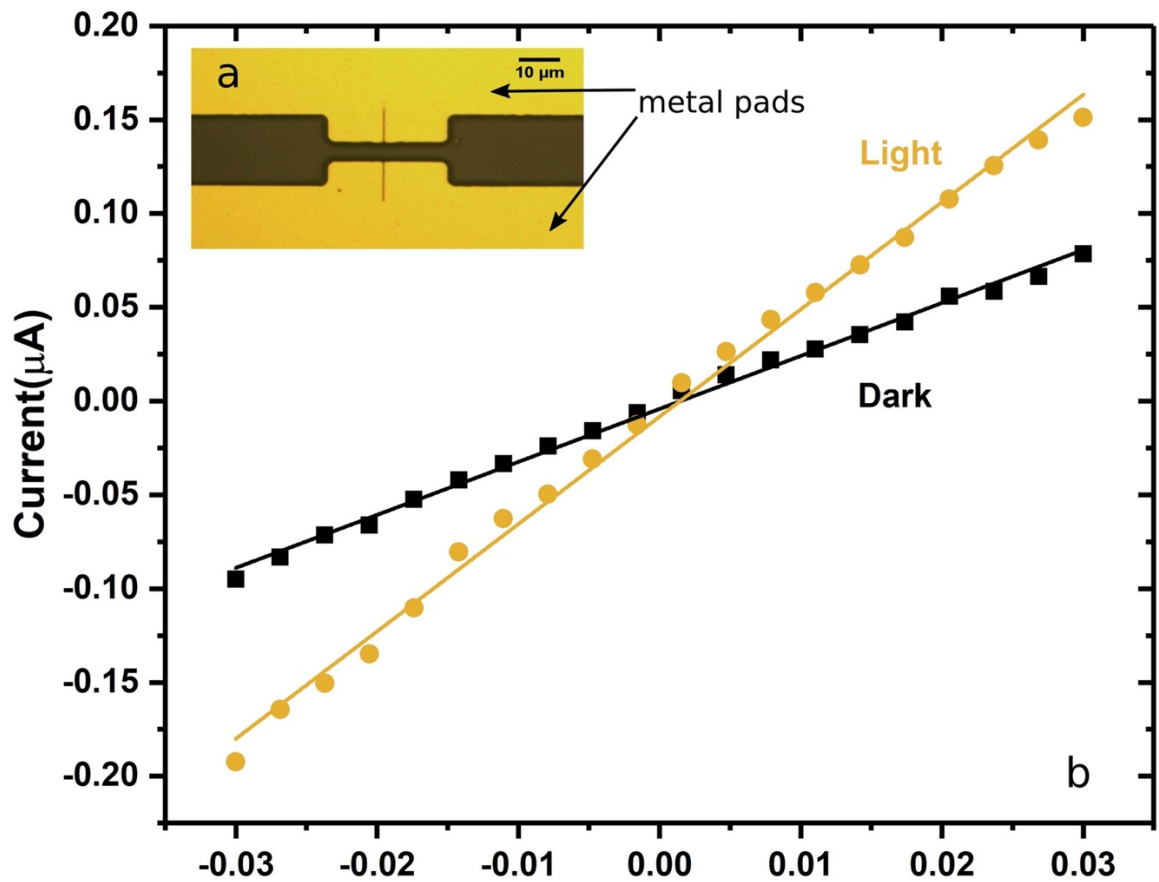

*Figure 2.IV characteristics of Wurtzite InAs nanowire device obtained in the dark and under white light illumination at 300K. Inset shows optical image of the nanowire device with Ti/Al pads.*

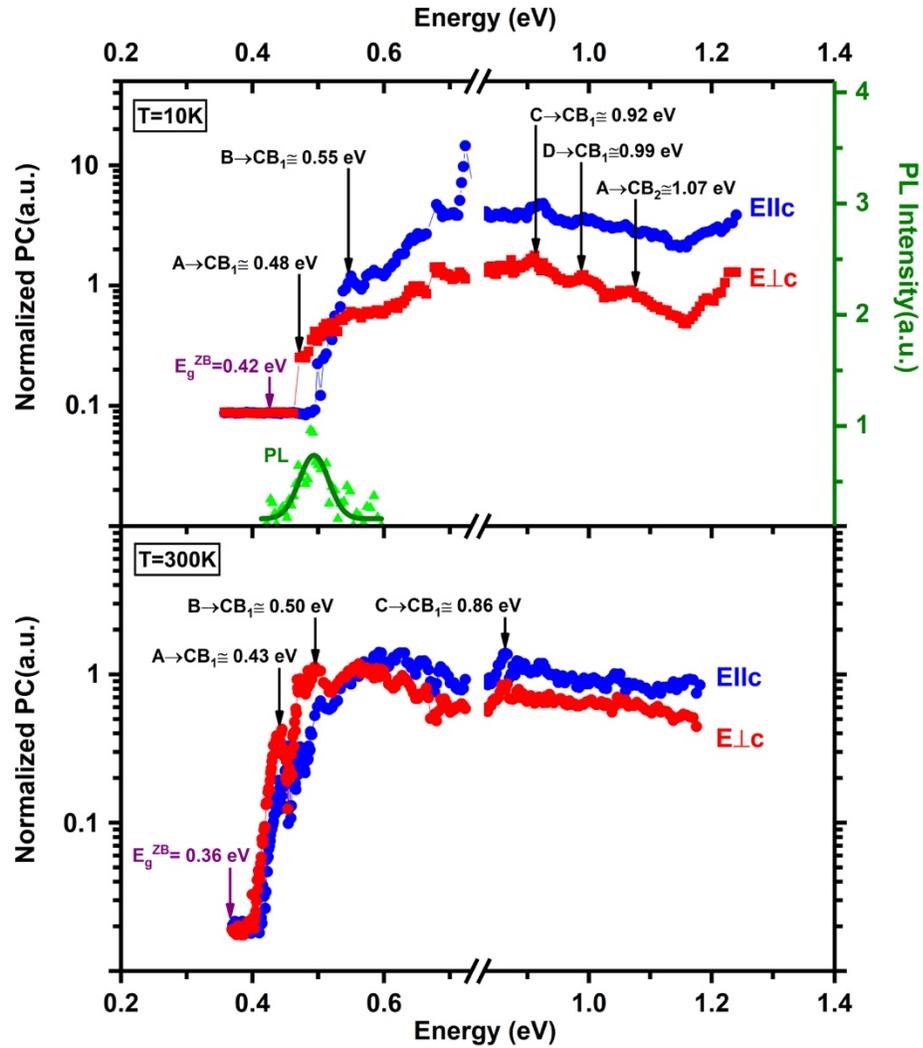

Figure 3. The photocurrent spectra *of a Wurtzite InAs nanowire acquired by exciting the carriers with polarization perpendicular to the nanowire (red squares) and parallel to the nanowire (blue circles) at temperature 300 K (on the top) and 10 K (bottom). The unpolarized PL spectra taken from a nanowire mechanically transferred on a Si substrate, shown by the green triangles and the fit solid line, confirms the fundamental bandgap at 10K. The energies of the peaks representing different excitations are marked on the diagram. The peaks of the higher temperature measurement are shifted to lower energies in comparison with 10K measurement, due to increased thermal energy and higher interatomic spacing.*

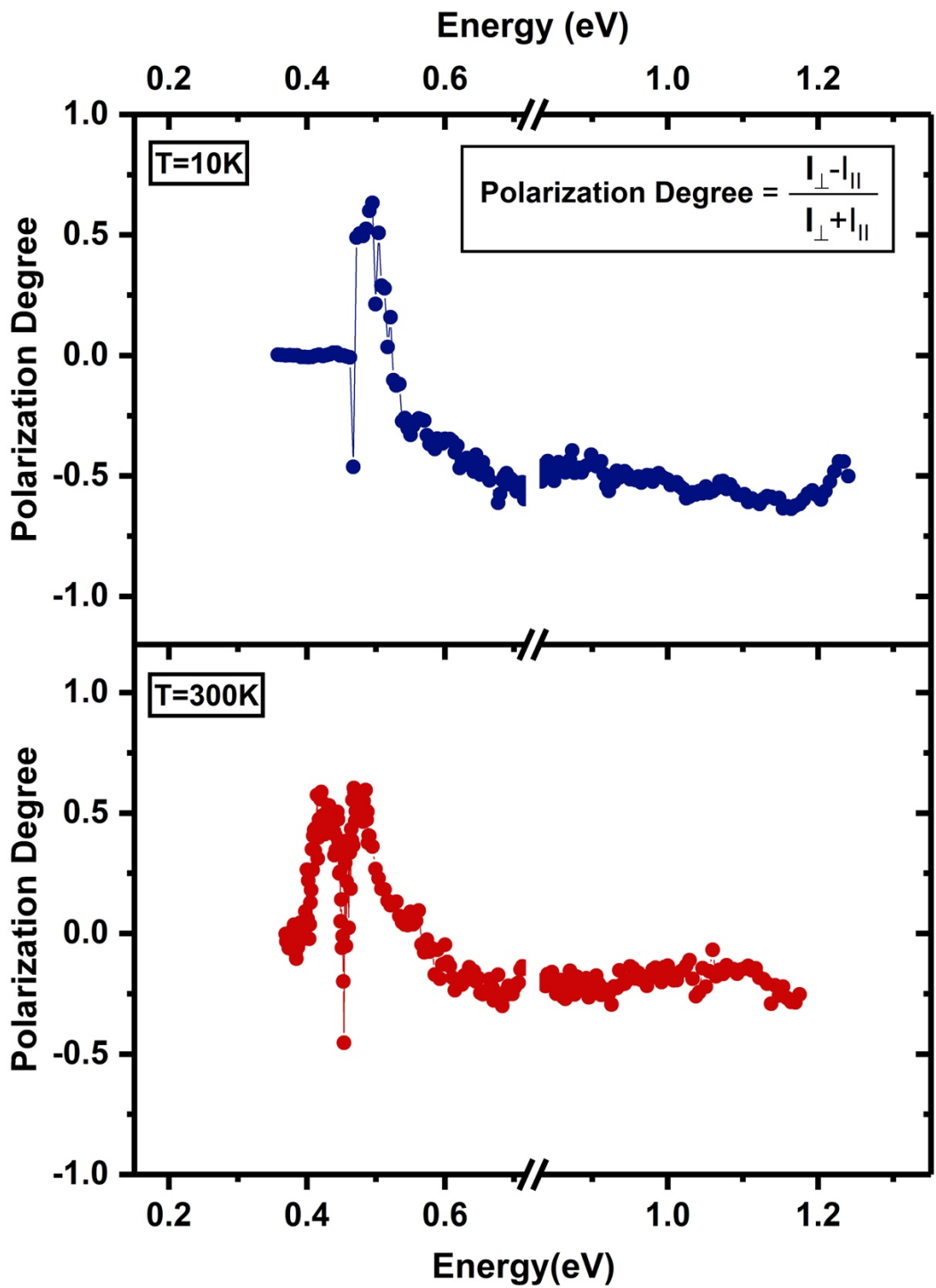

*Figure 4. Polarization of the photocurrent spectra at 10K and 300K calculated from data shown in Fig. 3.*

# Exploring the Band Structure of Wurtzite InAs Nanowires Using Photocurrent Spectroscopy

Seyyedesadaf Pournia, Samuel Linser, Giriraj Jnawali, Howard E. Jackson and Leigh M. Smith
*Department of Physics, University of Cincinnati, Cincinnati, OH 45221-0011*

Amira Ameruddin, Philippe Caroff, Jennifer Wong-Leung, Hark Hoe Tan, and Chennupati Jagadish
*Department of Electronic Materials Engineering, Research School of Physics, The Australian National University, Canberra, ACT 2601, Australia*

Hannah J. Joyce
*Department of Engineering, University of Cambridge, 9 JJ Thomson Avenue, Cambridge CB3 0FA, United Kingdom*

## Supplementary Information

Photocurrent measurements were taken repeatedly on different days to confirm that the peaks which we are considering as the transitions between the energy bands are reproducible. Some low temperature photocurrent spectra in the energy range of (0.82-1.24) eV are presented in Figure S1 below.

To identify each transition, we considered the selection rules determined from group theory (Equation 1 in the main text) and energetic order established from the theory in reference 1 to assign the peaks to the transitions. The energy values reported in Table S1 and also in the main text are the average of the energies in figure S1 and the errors are the standard deviations, which are summarized in the table below.

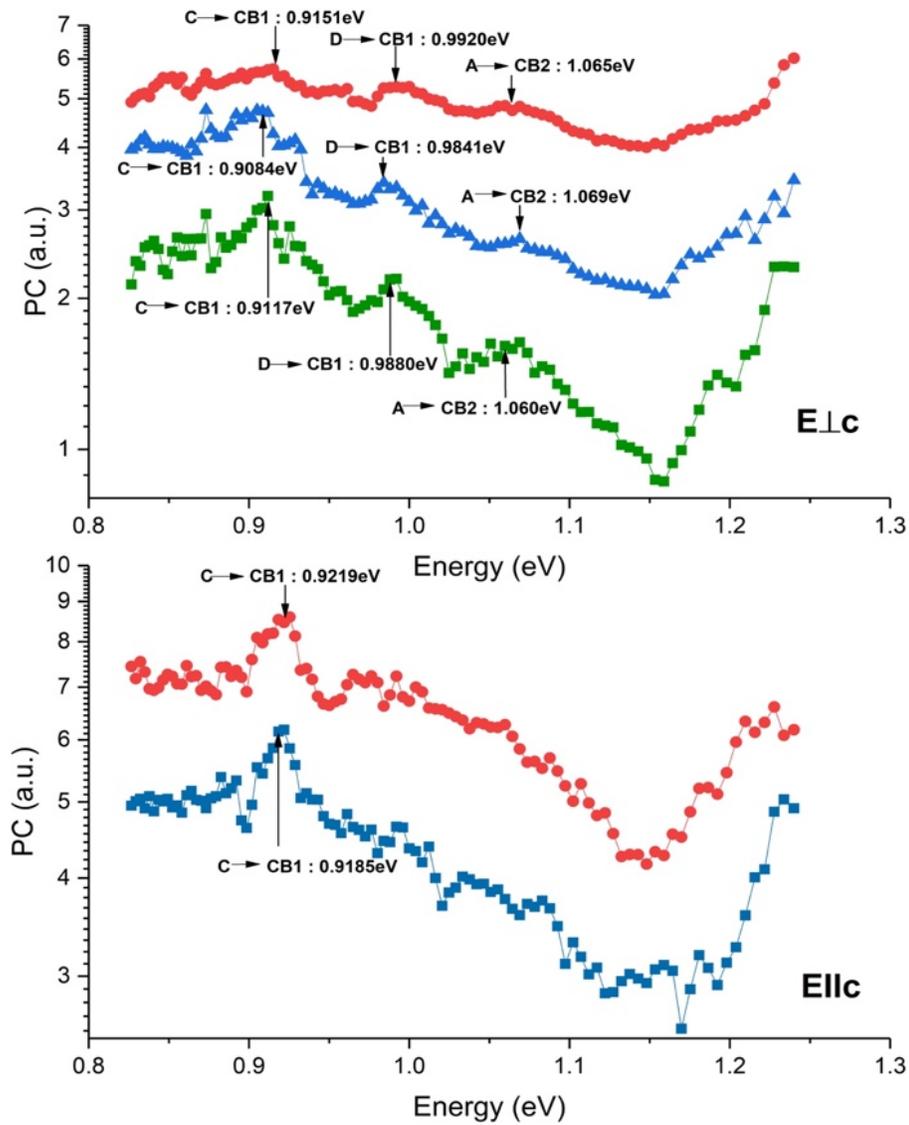

Figure S1. Photocurrent measurements on Wurtzite InAs device at T=10K with laser excitation polarized perpendicular (top) and parallel (bottom) to the c-axis of nanowire .

Table S1. Average energies and standard deviations.

| Transition | Mean (eV) | Standard Deviation (eV) |
|---|---|---|
| C→ CB1 | 0.915 | 0.005 |
| D→ CB1 | 0.988 | 0.003 |
| A→CB2 | 1.065 | 0.004 |